\documentstyle[preprint,aps,eqsecnum]{revtex}
\begin{document}
\tightenlines \draft

\newcommand{\beq}{\begin{equation}}
\newcommand{\eeq}{\end{equation}}
\newcommand{\bea}{\begin{eqnarray}}
\newcommand{\eea}{\end{eqnarray}}
\newcommand{\cir}{{\buildrel \circ \over =}}

\title{From Einstein's Hole Argument to Dirac and Bergmann
Observables.}

\medskip

\author{Luca Lusanna}

\address{
Sezione INFN di Firenze\\ Polo Scientifico\\ Via Sansone 1\\ 50019
Sesto Fiorentino (FI), Italy\\ E-mail LUSANNA@FI.INFN.IT}

\maketitle
\bigskip

Talk given at the 15th SIGRAV Conference on {\it General
Relativity and Gravitational Physics}, Villa Mondragone (Roma),
September 6-10, 2002.

\begin{abstract}

The individuation of point-events and the Hamiltonian way of
distinguishing gravitational from inertial effects in general
relativity are discussed.

\end{abstract}

\bigskip

The fact that both particle physics and all the approaches to
gravity make use of variational principles employing singular
Lagrangians \cite{1a} has the consequence that the Euler-Lagrange
equations cannot be put in normal form, some of them may be non
independent equations (due to the contracted Bianchi identities)
and a subset of the original configuration variables, the {\it
gauge variables}, are left completely undetermined. This leads to
the necessity of a division of the initial configuration variables
of any model in two groups:

i) the arbitrary {\it non determined gauge variables};

ii) the {\it gauge-invariant observables} with a deterministic
evolution.

But this process is in conflict with locality, manifest Lorentz
covariance, general covariance and, moreover, the configuration
space manifestly covariant approach has no natural analytical tool
to perform this separation.
\bigskip

Another non trivial aspect of the need of the division between
gauge variables and deterministic observables is the connection of
the latter with {\it measurable quantities}. Since, at least at
the classical level, the electro-magnetic measurable quantities
are the local electric and magnetic fields, we can extrapolate
that the {\it non-local} radiation gauge observables, i.e. the
transverse vector gauge potential and the transverse electric
field, are also measurable. But in the case of the non-Abelian
Yang-Mills gauge theories for the strong and weak interactions the
connection between gauge invariant observables and measurable
quantities is still poorly understood.

When we come to general relativity in Einstein formulation, these
problems become both more complex and more basic. More complex
because the Lie groups underlying the gauge groups of particle
physics are replaced by diffeomorphism groups
\footnote{Reparametrization invariant theories in Minkowski
space-time for particles and strings and parametrized Minkowski
theories for every isolated system \cite{1a,2a} also have
diffeomorphism groups as gauge groups.}, whose group manifold in
large is poorly understood. More basic because now the action of
the gauge group is not in an inner space of a field theory on a
background space-time, but is an extension to tensors over
space-time of the diffeomorphisms of the space-time itself. This
reflects itself in the much more singular nature of Einstein's
equations \footnote{Four of them are not independent from the
others due to the Bianchi identities, four are only restrictions
on the initial data and only two combinations of Einstein's
equations and their gradients depend on the {\it accelerations}
(the second time derivatives of the metric tensor).} with respect
to Yang-Mills equations. This fact has the dramatic consequence to
{\it destroy any physical individuality of the points of
space-time} as evidentiated by Einstein's {\it Hole Argument}
\cite{3a} in the years (1913-16) of the genesis of the concept of
general covariance. Only the idealization ({\it point-coincidence
argument}) according to which all possible observations reduce to
the intersections of the world-lines of observers, measuring
instruments and measured physical objects, convinced Einstein to
adopt general covariance and to abandon the physical objectivity
of space-time coordinates. In Einstein's own words\cite{3a}:

\begin{quotation}
{\footnotesize \noindent "That the requirement of general
covariance, {\it which takes away from space and time the last
remnant of physical objectivity}, is a natural one, will be seen
from the following reflexion. All our space-time verifications
invariably amount to a determination of space-time coincidences.
If, for example, events consisted merely in the motion of material
points, then ultimately nothing would be observable but the
meetings of two or more of these points. Moreover, the results of
our measurings are nothing but verifications of such meetings of
the material points of our measuring instruments with other
material points, coincidences between the hands of a clock and
points on the clock dial, and observed point-events happening at
the same place at the same time. The introduction of a system of
reference serves no other purpose than to facilitate the
description of the totality of such coincidences".}
\end{quotation}

\noindent At first sight it could seem  from these words that
Einstein simply equated general covariance with the unavoidable
{\it arbitrariness of the choice of coordinates}, a fact that, in
modern language, can be translated into {\it invariance under
passive diffeomorphisms}. Actually the essence of the {\it
point-coincidence argument} seems to be well in tune with the
Machian epistemology Einstein shared at the time, in particular as
regards the ontological privilege of "{\it bodies}" or "{\it
fields}" versus "{\it space}".

The Hole Argument, after a long oblivion, was resurrected by
Stachel \cite{4a} and then by Norton \cite{5a} and others as a
basic problem \cite{6a} in our both ontological and physical
understanding of {\it space-time} in general relativity, which is
commonly thought to imply that space-time points have no {\it
intrinsic physical meaning} due to the general covariance of
Einstein's equations. This feature is implicitly described in
standard modern textbooks by the statement that solutions to the
Einstein's equations related by ({\it active}) diffeomorphisms
have physically identical properties. Such kind of equivalence,
which also embodies the modern understanding of Einstein's {\it
Hole Argument}, has been named as {\it Leibniz equivalence} in the
philosophical literature by Earman and Norton \cite{7a} and
exploited to the effect of arguing against the {\it manifold
substantivalism} (it is not possible to reconcile the need for
determinism in classical physical laws and a {\it realistic}
interpretation of the mathematical 4-manifold describing
space-time) and in defense of the {\it relational} conception of
space-time. However {\it Leibniz equivalence} is not and cannot be
the last word about the {\it intrinsic physical properties} of
space-time, well beyond the needs of the empirical grounding of
the theory. In Ref.\cite{8a} (following Refs.\cite{9a}) an attempt
is made to gain an intrinsic {\it dynamical characterization of
space-time points in terms of the gravitational field itself},
besides and beyond the mathematical individuation furnished to
them by the coordinates. \bigskip

The basic mathematical concept that underlies the Hole Argument is
the concept of {\it active diffeomorphism} and its consequent
action on the tensor fields defined on a differentiable manifold.
Our manifold will be the mathematical manifold $M^4$, the first
layer of the would-be physical space-time of general relativity.
Consider a (geometrical or {\it active}) diffeomorphism $\phi$
which maps points of $M^4$ to points of $M^4$: $\phi: p
{\rightarrow \hspace{.2cm}}\ p' = \phi \cdot p$, and its tangent
map $\phi^{*}$ which maps tensor fields $T {\rightarrow
\hspace{.2cm}} \phi^{*} \cdot T$ in such a way that $[T](p)
{\rightarrow \hspace{.2cm}} [\phi^{*} \cdot T](p) \equiv
[T^{'}](p)$. Then $[\phi^{*} \cdot T](p) = [T](\phi^{-1}\cdot p)$.
It is seen that the transformed tensor field $\phi^{*} \cdot T$ is
a {\it new} tensor field whose components in general will have at
$p$ values that are {\it different} from those of the components
of $T$. On the other hand, the components of $\phi^* \cdot T$ have
at $p'$ - by construction -  the same values that the components
of the original tensor field $T$ have at $p$: $T^{'}(\phi \cdot p)
= T(p)$ or $T'(p) = T(\phi^{-1}\cdot p)$. The new tensor field
$\phi^* \cdot T$ is called the {\it drag-along} of $T$. Let us
recall that there is another, non-geometrical - so-called {\it
dual} - way of looking at the active diffeomorphisms, which,
incidentally, is more or less the way in which Einstein himself
formulated the original Hole Argument. This {\it duality} is based
on the circumstance that in each region of $M^4$ covered by two or
more charts there is a one-to-one correspondence between an {\it
active} diffeomorpshism and a specific coordinate transformation
(or {\it passive} diffeomorphism). The coordinate transformation
${\cal T}_{\phi}: x(p) {\rightarrow \hspace{.2cm}}\ x'(p) = [{\cal
T}_{\phi}x](p)$ which is {\it dual} to the active diffeomorphism
$\phi$ is defined such that $[{\cal T}_{\phi}x](\phi \cdot p) =
x(p)$. In its essence, this {\it duality} transfers the functional
dependence of the new tensor field in the new coordinate system to
the old system of coordinates. By analogy, the coordinates of the
new system $[x']$ are said to have been {\it dragged-along} with
the {\it active} diffeomorphism $\phi$.
\medskip

The right mathematical way of looking {\it passively} at the {\it
active} diffeomorphisms has been studied by  Bergmann  and Komar
\cite{10a}: they show that the group of active diffeomorphisms
(${}_ADiff\, M^4$) can be described by a non-normal sub-group of a
general group $Q$ of {\it passive dynamical symmetries} of
Einstein's equations [they correspond to generalized
metric-dependent coordinate transformations $x^{{'}\, \mu} =
f^{\mu}(x, g)$]. Q also contains two other non-normal sub-groups:
the passive diffeomorphisms (or coordinate transformations,
${}_PDiff\, M^4$) and the set of those (either active or passive)
diffeomorphisms which are projectable to phase space $Q_{can}$
(they are interpretable as Hamiltonian gauge transformations
generated by the first class constraints). Since passive
diffeomorphisms play the role of Lagrangian gauge transformations,
{\it a complete Lagrangian gauge fixing amounts to a definite
choice of the coordinates  on $M^4$}, a choice which, on the other
hand, is necessary in order to explicitly solve the Einstein
partial differential equations. In modern terminology, general
covariance implies that {\it a physical solution of Einstein's
equations} properly corresponds to a {\it 4-geometry}, namely the
equivalence class of all the 4-metric tensors, solutions of the
equations, written in all possible 4-coordinate systems. This
equivalence class is usually represented by the quotient ${}^4Geom
= {}^4Riem / {}_PDiff\, M^4$, where ${}^4Riem$ denotes the space
of metric tensors solutions of Einstein's equations. Then, any two
{\it inequivalent} Einstein space-times are different
4-geometries. As discussed in Ref.\cite{8a} Leibniz equivalence of
metric tensors $g$ means that an {\it Einstein (or on-shell, or
dynamical) gravitational field} is an equivalence class of
solutions of Einstein's equation {\it modulo} the dynamical
symmetry transformations of ${}_ADiff\, M^4$. We also have
${}^4Geom = {}^4Riem / {}_PDiff\, M^4 = {}^4Riem / Q = {}^4Riem /
{}_ADiff\, M^4 = {}^4Riem / Q_{can}$. It is clear that a
parametrization of the 4-geometries should be grounded on the two
independent dynamical degrees of freedom of the gravitational
field. At the Hamiltonian level the canonical reduction is done
{\it off-shell} (i.e. not on the solution of Einstein's
equations): an {\it off-shell gravitational field} is an
equivalence class under Hamiltonian gauge transformations
containing many different 4-geometries and only the restriction to
the solutions of Einstein's equations identifies a unique on-shell
4-geometry.
\medskip

Now, the Hole Argument, in its modern version, runs as follows.
Consider a general-relativistic space-time, as specified by the
four-dimensional mathematical manifold $M^4$ and by a metric
tensor field $g$ which {\it represents at the same time the
chrono-geometrical and causal structure of space-time and the
potential for the gravitational field}. The metric $g$ is a
solution of the generally-covariant Einstein equations. If any
non-gravitational physical fields are present, they are
represented by tensor fields that are also dynamical fields, and
that appear as sources in the Einstein equations. Assume now that
$M^4$ contains a {\it Hole} $\mathcal{H}$: that is, an open region
where all the non-gravitational fields are zero. On $M^4$ we can
prescribe an {\it active} diffeomorphism $\phi$ that re-maps the
points inside $\mathcal{H}$, but blends smoothly into the identity
map outside $\mathcal{H}$ and on the boundary. Now, {\it just
because Einstein's equations are generally covariant} so that they
can be written down as {\it geometrical relations}, if $g$ is one
of their solutions, so is the {\it drag-along} field $g' = \phi^*
\cdot g$. By construction, for any point $p \in \mathcal{H}$ we
have (geometrically) $g'(\phi \cdot p) = g(p)$, but of course
$g'(p) \neq g(p)$ (also geometrically). Now, what is the correct
interpretation of the new field $g'$? Clearly, the transformation
entails an {\it active redistribution of the metric over the
points of the manifold}, so the crucial question is whether, to
what extent, and how the points of the manifold are primarily {\it
individuated}.
\medskip

In the mathematical literature about topological spaces, it is
always implicitly assumed that the entities of the set can be
distinguished and considered separately (provided the Hausdorff
conditions are satisfied), otherwise one could not even talk about
point mappings or homeomorphisms. It is well known, however, that
the points of a homogeneous space cannot have any intrinsic
individuality\footnote{ As Hermann Weyl \cite{11a} puts it:
''There is no distinguishing objective property by which one could
tell apart one point from all others in a homogeneous space: at
this level, fixation of a point is possible only by a {\it
demonstrative act} as indicated by terms like {\it this} and {\it
there}.''}. There is only one way to individuate points at the
mathematical level: namely by {\it coordinatization}, a procedure
that transfers the individuality of $4$-tuples of real numbers to
the elements of the topological set. Precisely, one introduces by
convention {\it a standard} coordinate system for the {\it primary
individuation} of the points (like the choice of {\it standards}
in metrology). Then, one can get as many different {\it names},
for what we consider the same primary individuation, as the
coordinate charts containing the point in the chosen atlas of the
manifold. Therefore,  all the relevant transformations operated on
the manifold $M^4$ (including {\it active} diffeomorphisms which
map points to points), even if viewed in purely geometrical terms,
{\it must} be realizable in terms of coordinate transformations.

\medskip

If one now thinks of the (mathematically individuated) points of
$\mathcal{H}$ as also {\it physically individuated}
spatio-temporal events even before the metric is defined, then $g$
and $g'$ must be regarded as {\it physically distinct} solutions
of the Einstein equations (after all, as already noted, $g'(p)
\neq g(p)$ at the {\it same} point $p$). This, however, is a
devastating conclusion for the causality of the theory, because it
implies that, even after we completely specify a physical solution
for the gravitational and non-gravitational fields outside the
Hole - for example, on a Cauchy surface for the initial value
problem - we are still {\it unable to predict uniquely the
physical solution within the Hole}. As said the escape from the
(mathematical) strictures of the Hole Argument, is to {\it deny
that diffeomorphically related mathematical solutions represent
physically distinct solutions}. With this assumption, {\it an
entire equivalence class of diffeomorphically related mathematical
solutions represents only one physical solution}.

\medskip

It is seen at this point that the conceptual content of general
covariance is far more deeper than the simple invariance under
arbitrary changes of coordinates. Stachel \cite{12a,13a}  has
given a very enlightening analysis of the meaning of general
covariance and of its relations with the Hole Argument. He
stresses that asserting that $g$ and $\phi^* \cdot g$ represent
{\it one and the same gravitational field} is to imply that {\it
the mathematical individuation of the points of the differentiable
manifold by their coordinates has no physical content until a
metric tensor is specified}. In particular, coordinates lose any
{\it physical significance whatsoever} \cite{5a}. Furthermore, as
Stachel emphasizes, if $g$ and $\phi^* \cdot g$ must represent the
same gravitational field, they cannot be physically
distinguishable in any way. So when we act on $g$ with an active
diffeomorphisms to create the drag-along field $\phi^* \cdot g$,
{\it no element of physical significance} can be left behind: in
particular, nothing that could identify a point $p$ of the
manifold as the {\it same} point of space-time for both $g$ and
$\phi^* \cdot g$. Instead, when $p$ is mapped onto $p' = \phi
\cdot p$, it {\it brings over its identity}, as specified by
$g'(p')= g(p)$. A further important point made by Stachel is that
simply because a theory has generally covariant equations, it does
not follow that the points of the underlying manifold must lack
any kind of physical individuation. Indeed, what really matters is
that there can be no \emph{non-dynamical individuating field} that
is specified \emph{independently} of the dynamical fields, and in
particular independently of the metric. If this was the case, a
\emph{relative} drag-along of the metric with respect to the
(supposedly) individuating field would be physically significant
and would generate an inescapable Hole problem. Thus, the absence
of any non-dynamical individuating field, as well as of any
dynamical individuating field independent of the metric, is the
crucial feature of the purely gravitational solutions of general
relativity as well as of the very {\it concept} of {\it general
covariance}. In the case of general relativity there is {\it no
non-dynamical individuating field} like the distribution of rods
and clocks in Minkowsky space-time, that can be specified
independently of the dynamical fields, in particular independently
of the metric. This conclusion led Stachel to the conviction that
space-time points {\it must} be {\it physically} individuated {\it
before} space-time itself acquires a physical bearing, and that
the metric itself plays the privileged role of {\it individuating
field}: a necessarily {\it unique role} in the case of space-time
{\it without matter}. More precisely, Stachel claimed that this
individuating role should be implemented by four invariant
functionals of the metric, already considered by Bergmann and
Komar \cite{14a}. However, he did not follow up on his suggestion.
\bigskip

It is essential to realize that the {\it Hole Argument} is {\it
inextricably entangled} with the initial value problem. Most
authors have implicitly adopted the Lagrangian approach, where the
Cauchy problem is in-tractable because of the non-hyperbolic
nature of Einstein's equations (see Ref.\cite{15a} for an updated
review). The constrained Hamiltonian approach is just the {\it
only} proper way to analyze the initial value problem of that
theory and to find the {\it deterministically predictable
observables} of general relativity. It is not by chance that the
modern treatment of the initial value problem within the
Lagrangian configurational approach \cite{15a} must in fact mimic
the Hamiltonian methods. Only in the Hamiltonian approach can we
isolate the {\it gauge variables}, which carry the descriptive
arbitrariness of the theory, from the {\it Dirac observables}
(DO), which are gauge invariant quantities providing a
coordinatization of the reduced phase space of general relativity,
and are subjected to hyperbolic (and therefore  "{\it causal}" in
the customary sense) evolution equations. In physics the {\it Hole
Argument} is considered an aspect of the fact that also Einstein's
theory is interpreted as a gauge theory. The Leibnitz equivalence
is nothing else than the selection of the gauge invariant
observables of the theory. But now, differently from Yang-Mills
theories, the physical interpretation of the underlying
mathematical 4-manifold is lost, and this suggests that a
different interpretation of the gauge variables of generally
covariant theories with respect to Yang-Mills theories is needed.

\bigskip

As already said the manifestly covariant configuration space
approach has no natural tool to make a clean separation between
gauge variables and a basis of gauge invariant (hopefully
measurable) observables. Instead, at least locally, the
Hamiltonian formulation has natural tools for it, namely the
Shanmugadhasan canonical transformations \cite{16a}. The singular
Lagrangians of particle physics and general relativity imply the
use of Dirac-Bergmann theory \cite{17a,1a} of Hamiltonian
constraints  and only the constraint sub-manifold of phase space
is relevant for physics. Let us consider a finite-dimensional
system with configuration space $Q$ with global coordinates $q^i$,
$i=1,..,N$ described by a singular Lagrangian $L(q,\dot q)$
[${\dot q}^i(\tau ) = d\, q^i(\tau ) / d\tau$]. Let the Dirac
algorithm produce the following general pattern:

i) $m < N$ first class constraints $\phi_{\alpha}(q,p) \approx 0$,
of which the first $m_1 \leq m$ are primary, with the property
that the Poisson brackets of any two of them satisfies $\{
\phi_{\alpha}(q,p), \phi_{\beta}(q,p) \} =
C_{\alpha\beta\gamma}(q,p)\, \phi_{\gamma}(q,p) \approx 0$ ;

ii) $2n$ second class constraints, corresponding to pairs of
canonical variables which can be eliminated by going to Dirac
brackets;

iii) a Dirac Hamiltonian $H_D = H_c + \sum^m_{\alpha = m_1+1}\,
r_{\alpha}(q,p)\, \phi_{\alpha}(q,p) + \sum_{\alpha =1}^{m_1}\,
\lambda_{\alpha}(\tau )\, \phi_{\alpha}(q,p)$, where the
$\lambda_{\alpha}(\tau )$'s are arbitrary functions of time, named
{\it Dirac multipliers}, associated only with the {\it primary}
first class constraints \footnote{The use of the first half of
Hamilton equations, ${\dot q}^i = \{ q^i, H_D \}$, shows that the
Dirac multipliers are those primary velocity functions
($g_{\alpha}(q,\dot q) = \lambda_{\alpha}(\tau )$ on the solutions
of Hamilton equations) not determined by the singular
Euler-Lagrange equations. It can be shown that this arbitrariness
implies that also the secondary velocity functions
$r_{\alpha}(q,p) = {\tilde r}_{\alpha}(q,\dot q)$, $\alpha =
m_1+1,..,m$, in front of the secondary (and higher) first class
constraints in $H_D$, are not determined by the Euler-Lagrange
equations. Therefore each first class constraint has either a
configuration or a generalized velocity as an arbitrary partner.}.
In phase space there will be as many arbitrary Hamiltonian gauge
variables as first class constraints: they determine a
coordinatization of the {\it gauge orbits} inside the constraint
sub-manifold. The first class constraints are the generators of
the Hamiltonian gauge transformations under which the theory is
invariant and a gauge orbit is an equivalence class of all those
configurations which are connected by gauge transformations
(Leibnitz equivalence). The $2(N-m-n)$-dimensional reduced phase
space is obtained by eliminating the second class constraints with
Dirac brackets and by going to the quotient with respect to the
gauge orbits, or equivalently by adding as many gauge fixing
constraints as first class ones so to obtain $2m$ second class
constraints.

At least locally on the constraint sub-manifold the family of {\it
Shanmugadhasan canonical transformations} $q^i, p_i \mapsto
Q^{\alpha}, P_{\alpha} \approx 0, {\bar Q}^{\beta} \approx 0,
{\bar P}_{\beta} \approx 0, Q^A, P_A$, $\alpha =1,..,m$, $\beta =
1,..,n$, allows

i) to Abelianize the first class constraints, so that locally the
constraint submanifold is identified by the vanishing of a subset
of the new momenta $P_{\alpha} \approx 0$;

ii) to identify the associated Abelianized gauge variables
$Q^{\alpha}$ as coordinates parametrizing the gauge orbits;

iii) to replace the second class constraints with pairs of
canonical variables ${\bar Q}^{\beta} \approx 0$, ${\bar
P}_{\beta} \approx 0$;

iv) to identify a canonical basis of {\it gauge invariant Dirac
observables} with a deterministic evolution determined only by the
gauge invariant canonical part $H_c$ of the Dirac Hamiltonian.

This is the tool of the Hamiltonian formalism, lacking in the
configuration space approach, which allows to make the division
between arbitrary gauge variables and deterministic gauge
invariant observables. Since the (in general non local) Dirac
observables give a coordinatization of the classical reduced phase
space, it will depend on its topological properties whether a
given system with constraints admits a sub-family of
Shanmugadhasan canonical transformations globally defined. When
this happens the system admits {\it preferred} global separations
between gauge and observable degrees of freedom.

\bigskip

In ADM canonical gravity \cite{18a} \footnote{The existence of a
mathematical 4-manifold, the space-time $M^4$, admitting 3+1
splittings with space-like leaves $\Sigma_{\tau}\approx R^3$ is
assumed. All fields (also matter fields when present) depend on
$\Sigma_{\tau}$-adapted coordinates $(\tau ,\vec \sigma )$ for
$M^4$.} the ten components ${}^4g_{\mu\nu}$ of the 4-metric tensor
are replaced by the following configuration variables: the {\it
lapse} $N(\tau , \vec \sigma )$ and {\it shift} $N_r(\tau ,\vec
\sigma )$ functions and the six components of the {\it 3-metric
tensor} on $\Sigma_{\tau}$, ${}^3g_{rs}(\tau , \vec \sigma )$.
Einstein's equations are then recovered as the Euler-Lagrange
equations of the ADM action $S_{ADM} = - \epsilon k\int_{\triangle
\tau}d\tau \, \int d^3\sigma \, \lbrace \sqrt{\gamma} N\,
[{}^3R+{}^3K_{rs}\, {}^3K^{rs}-({}^3K)^2]\rbrace (\tau ,\vec
\sigma )$, which differs from Einstein-Hilbert action by a
suitable surface term. Here $k = {{c^3}\over {16 \pi\, G}}$,
${}^3K_{rs}$ is the extrinsic curvature of $\Sigma_{\tau}$,
${}^3K$ its trace, and ${}^3R$ the 3-curvature scalar. Besides the
ten configuration variables listed above, the ADM phase space is
{\it coordinatized} by ten canonical momenta ${\tilde \pi}^N(\tau
,\vec \sigma )$, ${\tilde \pi}^r_{\vec N}(\tau ,\vec \sigma )$,
${}^3{\tilde \Pi}^{rs}(\tau , \vec \sigma )$ \footnote{As shown in
Ref.\cite{19a}, a consistent treatment of the boundary conditions
at spatial infinity requires the explicit separation of the {\it
asymptotic} part of the lapse and shift functions from their {\it
bulk} part: $N(\tau ,\vec \sigma ) = N_{(as)}(\tau ,\vec \sigma )
+ n(\tau , \vec \sigma )$, $N_r(\tau ,\vec \sigma ) =
N_{(as)r}(\tau ,\vec \sigma ) + n_r(\tau , \vec \sigma )$, with
$n$ and $n_r$ tending to zero at spatial infinity in a
direction-independent way. On the contrary, $N_{(as)}(\tau ,\vec
\sigma ) = - \lambda_{\tau}(\tau ) - {1\over 2}\, \lambda_{\tau
u}(\tau )\, \sigma^u$ and $N_{(as)r}(\tau ,\vec \sigma ) = -
\lambda_{r}(\tau ) - {1\over 2}\, \lambda_{r u}(\tau )\,
\sigma^u$. The {\it Christodoulou-Klainermann space-times}
\cite{20a}, with their {\it rest-frame} condition of zero ADM
3-momentum and absence of super-translations, are singled out by
these considerations. The allowed foliations of these space-times
tend asymptotically to Minkowski hyper-planes in a
direction-independent way and are asymptotically orthogonal to the
ADM four-momentum. They have $N_{(as)}(\tau ,\vec \sigma ) =
\epsilon$, $N_{(as) r}(\tau ,\vec \sigma ) = 0$. Therefore, in
these space-times there are {\it asymptotic inertial time-like
observers} (the {\it fixed stars} or the {\it CMB rest frame}) and
the global mathematical time labeling the Cauchy surfaces can be
identified with their rest time. For the sake of simplicity these
aspects of the theory will be ignored, with the caveat that the
canonical pairs $N$, ${\tilde \pi}^N$, $N_r$, ${\tilde
\pi}^r_{\vec N}$ should be always replaced by the pairs $n$,
${\tilde \pi}^n$, $n_r$, ${\tilde \pi}^r_{\vec n}$. }. Such
canonical variables, however, are not independent since they are
restricted to the {\it constraint sub-manifold} by the eight {\it
first class} constraints

\bea
 {\tilde \pi}^N(\tau ,\vec \sigma ) &\approx& 0 ,\qquad
 {\tilde \pi}^r_{\vec N}(\tau ,\vec \sigma ) \approx 0,\nonumber \\
 &&{}\nonumber \\
 {\tilde {\cal H}}(\tau ,\vec \sigma )&=&\epsilon [k\sqrt{\gamma}\,
{}^3R-{1\over {2k \sqrt{\gamma}}} {}^3G_{rsuv}\, {}^3{\tilde
\Pi}^{rs}\, {}^3{\tilde \Pi}^{uv}] (\tau ,\vec \sigma ) \approx
0,\nonumber \\
 {}^3{\tilde {\cal H}}^r(\tau ,\vec \sigma )&=&-2\, {}^3{\tilde
\Pi}^{rs}{}_{| s} (\tau ,\vec \sigma )=-2[\partial_s\, {}^3{\tilde
\Pi}^{rs}+{}^3\Gamma^r_{su} {}^3{\tilde \Pi}^{su}](\tau ,\vec
\sigma ) \approx 0.
 \label{1}
  \eea

While the first four are {\it primary} constraints, the remaining
four are the super-hamiltonian and super-momentum  {\it secondary}
constraints arising from the requirement that the primary
constraints be constant in $\tau$. More precisely, this
requirement guarantees that, once we have chosen the initial data
inside the constraint sub-manifold  corresponding to a given
initial Cauchy surface $\Sigma_{\tau_o}$, the time evolution does
not take them out of the constraint sub-manifolds  for $\tau
> \tau_o$. The eight infinitesimal off-shell Hamiltonian gauge
transformations, generated by the first class constraints
(\ref{1}), have the following interpretation\cite{19a}:

i) those generated by the four primary constraints modify the
lapse and shift functions: these in turn determine how densely the
space-like hyper-surfaces $\Sigma_{\tau}$ are distributed in
space-time and also the conventions to be made on each
$\Sigma_{\tau}$ about simultaneity (the choice of clocks
synchronization) and gravito-magnetism;

ii) those generated by the three super-momentum constraints induce
a transition on $\Sigma_{\tau}$ from a given 3-coordinate system
to another one;

iii) that generated by the super-hamiltonian constraint induces a
transition from a given 3+1 splitting of $M^4$ to another one, by
operating normal deformations of the space-like
hyper-surfaces\footnote{Note that in {\it compact} space-times the
super-hamiltonian constraint is usually interpreted as generator
of the evolution in some {\it internal time}, either like York's
internal {\it extrinsic} time or like Misner's internal {\it
intrinsic} time. Here instead the super-hamiltonian constraint is
the generator of those Hamiltonian gauge transformations which
imply that the description is independent of the choice of the
allowed 3+1 splitting of space-time: {\it this is the correct
answer to the criticisms raised against the phase space approach
on the basis of its lack of manifest covariance}. }.
\bigskip

The evolution in $\tau$ is ruled by the Hamilton-Dirac Hamiltonian

\beq
 H_{(D)ADM}=\int d^3\sigma \, \Big[ N\, {\tilde {\cal H}}+N_r\,
{}^3{\tilde {\cal H}}^r + \lambda_N\, {\tilde \pi}^N + \lambda
^{\vec N}_r\, {\tilde \pi}^r_{\vec N}\Big](\tau ,\vec \sigma )
\approx 0,
 \label{2}
 \eeq

\noindent where $\lambda_N(\tau , \vec \sigma )$ and
$\lambda^r_{\vec N}(\tau ,\vec \sigma )$ are {\it arbitrary Dirac
multipliers} in front of the primary constraints\footnote{These
are four {\it velocity functions} (gradients of the metric tensor)
which are not determined by Einstein's equations. As shown in
Ref.\cite{19a}, the correct treatment of the boundary conditions
leads to rewrite Eq.(\ref{2}) in terms of $n$ and $n_r$, which are
the arbitrary secondary velocity functions.}. This is just the
Hamiltonian counterpart of the so-called "{\it indeterminism}"
surfacing in the Hole Argument. The resulting hyperbolic system of
Hamilton-Dirac equations has the same solutions of the
non-hyperbolic system of (Lagrangian) Einstein's equations with
the same boundary conditions.

\medskip

The off-shell freedom corresponding to the eight independent types
of Hamiltonian gauge transformations is reduced on-shell to four
types like in the case of ${}_PDiff\, M^4$: precisely the
transformations in [$Q_{can} \cap {}_PDiff\, M^4$] . At the
off-shell level, this property is manifest by the circumstance
that the original Dirac Hamiltonian contains {\it only} 4
arbitrary Dirac multipliers and that the {\it correct gauge-fixing
procedure} \cite{19a} starts by giving {\it only} the four gauge
fixing constraints for the secondary constraints. The requirement
of time constancy then generates the four gauge fixing constraints
to the primary constraints, while time constancy of such secondary
gauge fixings leads to the determination of the four Dirac
multipliers. Since the original constraints plus the above eight
gauge fixing constraints form a second class set, it is possible
to introduce the associated {\it Dirac brackets} and conclude the
canonical reduction by realizing an off-shell reduced phase space.
Of course, once a {\it completely fixed Hamiltonian gauge} is
reached, general covariance is completely broken. Note that a
completely fixed Hamiltonian gauge {\it on-shell} is equivalent to
a {\it definite choice of the space-time 4-coordinates} on $M^4$
within the Lagrangian viewpoint.

\bigskip

In order to visualize the meaning of the various types of degrees
of freedom  one needs a determination of a {\it Shanmugadhasan
canonical basis} \cite{16a} of metric gravity \cite{19a} having
the following structure ($\bar a =1,2$ are non-tensorial indices
of the DO \footnote{The DO are in general neither tensors nor
invariants under space-time diffeomorphisms. Therefore their
(unknown) functional dependence on the original variables changes
(off-shell) with the gauge and, therefore, (on-shell) with the
4-coordinate system.} $r_{\bar a}$, $\pi_{\bar a}$) with

 \bea
\begin{minipage}[t]{3cm}
\begin{tabular}{|l|l|l|} \hline
$N$ & $N_r$ & ${}^3g_{rs}$ \\ \hline ${\tilde \pi}^N \approx 0$ &
${\tilde \pi}_{\vec N}^r \approx 0$ & ${}^3{\tilde \Pi}^{rs}$ \\
\hline
\end{tabular}
\end{minipage} &&\hspace{2cm} {\longrightarrow \hspace{.2cm}} \
\begin{minipage}[t]{4 cm}
\begin{tabular}{|ll|l|l|l|} \hline
$N$ & $N_r$ & $\xi^{r}$ & $\phi$ & $r_{\bar a}$\\ \hline
 ${\tilde \pi}^N \approx 0$ & ${\tilde \pi}_{\vec N}^r \approx 0$
& ${\tilde \pi}^{{\vec {\cal H}}}_r \approx 0$ &
 $\pi_{\phi}$ & $\pi_{\bar a}$ \\ \hline
\end{tabular}
\end{minipage} \nonumber \\
 &&{}\nonumber \\
&& {\longrightarrow \hspace{.2cm}} \
\begin{minipage}[t]{4 cm}
\begin{tabular}{|ll|l|l|l|} \hline
$N$ & $N_r$ & $\xi^{r}$ & $Q_{\cal H} \approx 0$ & $r^{'}_{\bar
a}$\\ \hline
 ${\tilde \pi}^N \approx 0$ & ${\tilde \pi}^r_{\vec N} \approx 0$
& ${\tilde \pi}^{{\vec {\cal H}}}_r \approx 0$ &
 $\Pi_{\cal H}$ & $\pi^{'}_{\bar a}$ \\ \hline
\end{tabular}
\end{minipage}.
 \label{3}
 \eea
\medskip

\noindent It is seen that we need a sequence of two canonical
transformations.

a) The first one replaces seven first-class constraints with as
many Abelian momenta ($\xi^r$ are the gauge parameters of the
passive 3-diffeomorphisms generated by the super-momentum
constraints) and introduces the conformal factor $\phi$ of the
3-metric as the configuration variable to be determined by the
super-hamiltonian constraint  \footnote{Recall that the {\it
strong} ADM energy is the flux through the surface at spatial
infinity of a function of the 3-metric only, and it is weakly
equal to the {\it weak} ADM energy (volume form) which contains
all the dependence on the ADM momenta. This implies \cite{19a}
that the super-hamiltonian constraint must be interpreted as the
equation ({\it Lichnerowicz equation}) that uniquely determines
the {\it conformal factor} $\phi = ( det\, {}^3g )^{1/12}$ of the
3-metric as a functional of the other variables. This means that
the associated gauge variable is the {\it canonical momentum
$\pi_{\phi}$ conjugate to the conformal factor}: this latter
carries information about the extrinsic curvature of
$\Sigma_{\tau}$. It is just this variable, and {\it not} York's
time, which parametrizes the {\it normal} deformation of the
embeddable space-like hyper-surfaces $\Sigma_{\tau}$.}. Note that
the final gauge variable, namely the momentum $\pi_{\phi}$
conjugate to the conformal factor, is the only gauge variable of
momentum type: it plays the role of a {\it time} variable, so that
the Lorentz signature of space-time is made manifest by the
Shanmugadhasan transformation in the set of gauge variables
$(\pi_{\phi}; \xi^r)$. More precisely, the first canonical
transformation should be called a {\it quasi-Shanmugadhasan }
transformation, because nobody has succeeded so far in
Abelianizing the super-hamiltonian constraint. Note furthermore
that this transformation is a {\it point} canonical
transformation.

b) The second canonical transformation would be instead a {\it
complete Shanmugadhasan} transformation, where $Q_{{\cal H}}(\tau
,\vec \sigma ) \approx 0$ would denote the Abelianization of the
super-hamiltonian constraint \footnote{If $\tilde \phi [r_{\bar
a}, \pi_{\bar a}, \xi^r, \pi_{\phi}]$ is the solution of the
Lichnerowicz equation, then $Q_{{\cal H}}=\phi - \tilde \phi
\approx 0$. Other forms of this canonical transformation should
correspond to the extension of the York map \cite{21a} to
asymptotically flat space-times: in this case the momentum
conjugate to the conformal factor would be just York time and one
could add the maximal slicing condition as a gauge fixing. Again,
however, nobody has been able so far to build a York map
explicitly.}. The variables $N$, $N_r$, $\xi^r$, $\Pi_{\cal H}$
are the final {\it Abelianized Hamiltonian gauge variables} and
$r^{'}_{\bar a}$, $\pi^{'}_{\bar a}$ the final DO. In absence of
explicit solutions of the Lichnerowicz equation, the best we can
do is to construct the {\it quasi-Shanmugadhasan} transformation.
On the other hand, such transformation has the remarkable property
that, in the {\it special gauges} with $\pi_{\phi}(\tau ,\vec
\sigma ) \approx 0$, the variables $r_{\bar a}$, $\pi_{\bar a}$
form a canonical basis of off-shell DO for the gravitational field
{\it even if} the solution of the Lichnerowicz equation is not
known.

\bigskip

The four gauge fixings to the secondary constraints, when written
in the quasi-Shanmugadhasan canonical basis, have the following
meaning:

 i) the three gauge fixings for the parameters $\xi^r$ of the
spatial passive diffeomorphisms generated by the super-momentum
constraints correspond to the choice of a system of 3-coordinates
on $\Sigma_{\tau}$\footnote{Since the diffeomorphism group has no
canonical identity, this gauge fixing has to be done in the
following way. One chooses a 3-coordinate system by choosing a
parametrization of the six components ${}^3g_{rs}(\tau ,\vec
\sigma )$ of the 3-metric in terms of {\it only three} independent
functions. This amounts to fix the three functional degrees of
freedom associated with the diffeomorphism parameters $\xi^r(\tau
,\vec \sigma )$. For instance, a 3-orthogonal coordinate system is
identified by ${}^3g_{rs}(\tau ,\vec \sigma ) = 0$ for $r \not= s$
and ${}^3g_{rr} = \phi^2\,  exp(\sum_{\bar a = 1}^2 \gamma_{r\bar
a} r_{\bar a})$. Then, one imposes the gauge fixing constraints
$\xi^r(\tau ,\vec \sigma ) - \sigma^r \approx 0$ as a way of
identifying this system of 3-coordinates with a conventional
origin of the diffeomorphism group manifold. }. The time constancy
of these gauge fixings generates the gauge fixings for the shift
functions $N_r$ while the time constancy of the latter leads to
the fixation of the Dirac multipliers $\lambda^{\vec N}_r$;

ii) the gauge fixing to the super-hamiltonian constraint
determines $\pi_{\phi}$: it is a fixation of the form of
$\Sigma_{\tau}$ and amounts to the choice of one particular 3+1
splitting of $M^4$. Since the time constancy of the gauge fixing
on $\pi_{\phi}$ determines the gauge fixing for the lapse function
$N$ (and then of the Dirac multiplier $\lambda_N$), it follows a
connection with the choice of the standard of local proper time.

All this entails that, after such a fixation of the gauge $G$, the
functional form of the DO in terms of the original variables
becomes gauge-dependent. At this point it is convenient to denote
them as $r^G_{\bar a}$, $\pi^G_{\bar a}$. Since the Shanmugadhasan
canonical transformation is a {\it highly non-local}
transformation and it is not known how to build a global atlas of
coordinate charts for the group manifold of diffeomorphism groups,
it is not known either how to express the $\xi^r$'s, $\pi_{\phi}$
and the DO in terms of the original ADM canonical
variables.\footnote{This should be compared to the Yang-Mills
theory in case of a trivial principal bundle, where the
corresponding variables are defined by a path integral over the
original canonical variables \cite{22a,1a}.}. In conclusion, a
representative of a {\it Hamiltonian kinematic or off-shell
gravitational field}, in a given gauge equivalence class, is
parametrized by $r_{\bar a}$, $\pi_{\bar a}$ and is an element of
a {\it conformal gauge orbit} (it contains all the 3-metrics in a
conformal 3-geometry) spanned by the gauge variables $\xi^r$,
$\pi_{\phi}$, $N$, $N_r$. Therefore, according to the gauge
interpretation based on constraint theory, a {\it Hamiltonian
kinematic or off-shell gravitational field} is an equivalence
class of 4-metrics modulo the Hamiltonian group of gauge
transformations, which contains a well defined conformal
3-geometry.
\bigskip

The previous discussion applies to a class of globally hyperbolic,
topologically trivial, non-compact (asymptotically flat at spatial
infinity) space-times of the type of Christodoulou-Klainermann
ones \cite{20a}. In them we have \cite{19a,8a}:
\medskip

1) The imposition of suitable boundary conditions on the fields
and the gauge transformations of canonical ADM metric gravity
eliminates the super-translations and reduces the asymptotic
symmetries at spatial infinity to the asymptotic ADM Poincar\'e
group. The asymptotic implementation of Poincar\'e group makes
possible the general-relativistic definition of angular momentum
and the matching of general relativity with particle physics.

2) The boundary conditions of point 1) require that the leaves of
the foliations associated with the admissible 3+1 splittings of
space-time must tend to Minkowski space-like hyper-planes
asymptotically orthogonal to the ADM 4-momentum in a
direction-independent way. This property is concretely enforced by
using a technique introduced by Dirac \cite{16a} for the selection
of space-times admitting asymptotically flat 4-coordinates at
spatial infinity. \footnote{ Dirac's method brings to an
enlargement of ADM canonical metric gravity with non-vanishing ADM
Poincar\'e charges. Such space-times admit preferred asymptotic
inertial observers, interpretable as fixed stars (the standard for
measuring rotations). Such non-Machian properties allow to merge
the standard model of elementary particles in general relativity
with all the (gravitational and non-gravitational) fields
belonging to the same function space (suitable weighted Sobolev
spaces). Besides the existence of a realization of the Poincar\'e
group, only one additional property is required: namely that the
space-like hyper-surfaces admit an involution \cite{23a} allowing
the definition of a generalized Fourier transform with its
associated concepts of positive and negative energy. This
disproves the claimed impossibility of defining particles in
curved space-times \cite{24a}.}

3) The super-hamiltonian constraint is the generator of the gauge
transformations connecting different admissible 3+1 splittings of
space-time and {\it has nothing to do with the temporal
evolution}.

4) As shown by DeWitt \cite{25a},  the weakly vanishing ADM Dirac
Hamiltonian has to be modified with a suitable surface term in
order that functional derivatives, Poisson brackets and Hamilton
equations be mathematically well- defined in such non-compact
space-times. This fact, in conjunction with the points 1), 2), 3)
above, entails that {\it there is an effective evolution in the
mathematical time} $\tau$ which parametrizes the leaves of the
foliation associated with any 3+1 splitting. Such evolution is
ruled by the weak {\it ADM energy} \cite{19a,26a}, i.e. by a
non-vanishing Hamiltonian which exists also in the reduced phase
space. This is the {\it rest-frame instant form of metric gravity}
\cite{19a}. Each gauge fixing creates a realization of the reduced
phase space and the {\it weak {\it ADM} energy} is a functional of
only the DO of that gauge. Then, the DO themselves (as any other
function of them) satisfy the Hamilton equations $\dot r^G_{\bar
a} = \{ r^G_{\bar a}, E_{\mathrm{ADM}}\}^*, \quad \dot \pi^G_{\bar
a} = \{\pi^G_{\bar a}, E_{\mathrm{ADM}}\}^*$, where
$E_{\mathrm{ADM}}$ is intended as the restriction of the weak {\it
ADM} Energy to the reduced phase space and where the
$\{\cdot,\cdot\}^*$ are Dirac Brackets.

\bigskip

5) When matter is present in this family of space-times, switching
off Newton's constant  ($G \mapsto 0$) yields the description of
matter in Minkowski space-time foliated with the space-like
hyper-planes orthogonal to the total matter 4-momentum (Wigner
hyper-planes intrinsically defined by matter isolated system). In
this way one gets the rest-frame instant form of dynamics
reachable from parametrized Minkowski theories \cite{1a}.
Incidentally, this is the first example of consistent
deparametrization of general relativity in which the ADM
Poincar\'e group tends to the Poincar\'e group of the isolated
matter system.

\bigskip

These space-times are a  {\it counterexample to the frozen time
argument} based on the widespread opinion (see for instance
Refs.\cite{27a}) that the Hamiltonian approach to general
relativity is not fruitful, because it leads to a reduced phase
space, which is a {\it frozen} space without evolution. For
instance Belot and Earman \cite{27a} draw ontological conclusions
about the absence of real (temporal) change in general relativity
from the circumstance that, in spatially compact models of general
relativity, the Hamiltonian temporal evolution boils down to a
mere gauge transformation and is, therefore, physically
meaningless. Instead in the previous space-times there is neither
a frozen reduced phase space nor a Wheeler-DeWitt interpretation
based on some local concept of time like in compact space-times.
Therefore, {\it our gauge-invariant approach to general relativity
is perfectly adequate to accommodate objective temporal change}.

\bigskip

In the previous  Hamiltonian context there are the tools for
completing Stachel's suggestion and exploiting the old proposal
advanced by Bergmann and Komar \cite{14a} for an intrinsic
labeling of space-time points by means of the eigenvalues of the
Weyl tensor. Its four invariant scalar eigenvalues
$\Lambda^{(k)}_W(\tau ,\vec \sigma )$, $k=1,..,4$,  written in
Petrov compressed notations, are $\Lambda^{(1)}_W = Tr\, ({}^4C\,
{}^4g\, {}^4C\, {}^4g)$, $\Lambda^{(2)}_W = Tr\, ({}^4C\, {}^4g\,
{}^4C\, {}^4\epsilon )$, $\Lambda^{(3)}_W = Tr\, ({}^4C\, {}^4g\,
{}^4C\, {}^4g\, {}^4C\, {}^4g)$, $\Lambda^{(4)}_W = Tr\, ({}^4C\,
{}^4g\, {}^4C\, {}^4g\, {}^4C\, {}^4\epsilon )$, where ${}^4C$ is
the Weyl tensor, $^4g$ the metric, and ${}^4\epsilon$ the
Levi-Civita totally anti-symmetric tensor. Bergman and Komar
\cite{14a,28a,29a} proposed that we build a set of (off-shell)
{\it invariant pseudo-coordinates} for the point-events of
space-time as four suitable functions of the $\Lambda^{(k)}_W$'s,
${\bar \sigma}^{\bar A}(\sigma ) = F^{\bar A}[\Lambda^{(k)}_W
 [{}^4g(\sigma), \partial {}^4g(\sigma)]],
\,(\bar A = 1,2,...,4)$. Indeed, under the hypothesis of no
space-time symmetries, we would be tempted (like Stachel) to use
the $F^{\bar A}[\Lambda^{(k)}_W]$ as individuating fields to {\it
label the points of space-time}, at least locally. Of course,
since they are invariant functionals, the $F^{\bar
A}[\Lambda^{(k)}_W]$'s are quantities invariant under passive
diffeomorphisms (PDIQ), therefore, as such, they do not define a
coordinate chart for the atlas of the mathematical Riemannian
4-manifold $M^4$ in the usual sense (hence the name of {\it
pseudo-coordinates} and the superior bar  used in $F^{\bar A}$).
Moreover, the tetradic 4-metric which can be built by means of the
intrinsic pseudo-coordinates  is a formal object invariant under
passive diffeomorphisms that does not satisfy Einstein's equations
(but possibly much more complex derived equations).

\medskip

The procedure of point identification starts from the fact that,
within the Hamiltonian approach, Bergmann and Komar \cite{14a}
proved the fundamental result that the Weyl eigenvalues
$\Lambda^{(A)}_W$, once re-expressed as functionals of the Dirac
(i.e. ADM) canonical variables, {\it do not depend on the lapse
and shift functions} but only on the 3-metric and its conjugate
canonical momentum, $\Lambda^{(k)}_W[{}^4g(\tau ,\vec \sigma),
\partial {}^4g(\tau ,\vec \sigma)] = {\tilde \Lambda}^{(k)}_W[{}^3g(\tau
,\vec \sigma), {}^3\Pi(\tau ,\vec \sigma)]$. This result is
crucial since it entails that just the {\it intrinsic
pseudo-coordinates} ${\bar \sigma}^{\bar A}$ can be exploited as
natural and peculiar {\it coordinate gauge conditions} in the
canonical reduction procedure. In a completely fixed (either {\it
off}- or {\it on-shell}) gauge, both the four intrinsic {\it
pseudo-coordinates} and the ten {\it tetradic} components of the
metric field   become gauge dependent functions of the four DO of
that gauge. For the Weyl scalars in particular we can write
$\Lambda^{(k)}_W(\tau ,\vec \sigma ){|}_G =  {\tilde
\Lambda}^{(k)}_W[{}^3g(\tau ,\vec \sigma), {}^3\Pi(\tau ,\vec
\sigma)]{|}_G = \Lambda_G^{(k)}[r^G_{\bar a}(\tau ,\vec \sigma ),
\pi^G_{\bar a}(\tau ,\vec \sigma )]$, where ${|}_G$ denotes the
specific gauge. Conversely, by the inverse function theorem, in
each gauge, the DO of that gauge can be expressed as functions of
the 4 eigenvalues restricted to that gauge, $\Lambda^{(k)}_W (\tau
,\vec \sigma ){|}_G $.
\medskip

Bergmann-Komar proposal can be utilized in constructing a peculiar
{\it gauge-fixing to the super-hamiltonian and super-momentum
constraints} in the canonical reduction of general relativity in
the following way: after having selected a {\it completely
arbitrary mathematical} coordinate system $\sigma^A \equiv
[\tau,\sigma^a]$ adapted to the $\Sigma_\tau$ surfaces, one
chooses {\it as physical individuating fields} four suitable
functions $F^{\bar A}[\Lambda^{(k)}_W(\tau ,\vec \sigma )]$, and
expresses them as functionals $\tilde F^{\bar A}$ of the ADM
variables $F^{\bar A}[\Lambda^{(k)}_W(\tau ,\vec \sigma )] =
F^{\bar A}[{\tilde \Lambda}^{(k)}_W[{}^3g(\tau ,\vec \sigma),
{}^3\Pi(\tau ,\vec \sigma)]] = \tilde F^{\bar A}[{}^3g(\tau ,\vec
\sigma), {}^3\Pi(\tau ,\vec \sigma)]$. The space-time points, {\it
mathematically individuated} by the quadruples of real numbers
$\sigma^A$, become now {\it physically individuated point-events}
through the imposition of the following gauge fixings to the four
secondary constraints

\beq
 {\bar \chi}^A(\tau ,\vec \sigma )\- {\buildrel {def} \over =}\-  \sigma^A -
 {\bar \sigma}^{\bar A}(\tau ,\vec \sigma ) = \sigma^A -
 F^{\bar A}\Big[{\tilde \Lambda}^{(k)}_W[{}^3g(\tau ,\vec \sigma), {}^3\Pi(\tau ,\vec
\sigma)]\Big] \approx 0.
 \label{4}
 \eeq

\noindent Then, following the standard procedure a completely
fixed Hamiltonian gauge, say $G$, is defined. This will be a
correct gauge fixing provided the functions $F^{\bar A}[{
\Lambda}^{(k)}_W(\tau ,\vec \sigma )]$ are chosen so that the
${\bar \chi}^A(\tau ,\vec \sigma )$'s satisfy the {\it orbit
conditions} $det\, | \{ {\bar \chi}^A(\tau ,\vec \sigma ), {\tilde
{\cal H}}^B(\tau ,{\vec \sigma}^{'}) \} | \not= 0$, where ${\tilde
{\cal H}}^B(\tau ,\vec \sigma ) = \Big( {\tilde {\cal H}}(\tau
,\vec \sigma ); {}^3{\tilde {\cal H}}^r(\tau ,\vec \sigma ) \Big)
\approx 0$ are the super-hamiltonian and super-momentum
constraints of Eqs.(\ref{1}). These conditions enforce the Lorentz
signature on Eq.(\ref{4}), namely the requirement that $F^{\bar
\tau}$ be a {\it time} variable, and imply that {\it the $F^{\bar
A}$'s are not DO}. The above gauge fixings allow in turn the
determination of the four Hamiltonian gauge variables $\xi^r(\tau
,\vec \sigma )$, $\pi_{\phi}(\tau ,\vec \sigma )$ of
Eqs.(\ref{3}). Then, their time constancy induces the further
gauge fixings ${\bar \psi}^A(\tau ,\vec \sigma ) \approx 0$ for
the determination of the remaining gauge variables, i.e., the
lapse and shift functions in terms of the DO in that gauge as

\bea
 {\dot {\bar \chi}}^A(\tau ,\vec \sigma ) &=& {{\partial {\bar
 \chi}^A(\tau ,\vec \sigma )} \over {\partial \tau}} + \{
 {\bar \sigma}^{\bar A}(\tau ,\vec \sigma ), {\bar H}_D \} = \delta^{A\tau}
 +\nonumber \\
 &+& \int d^3\sigma_1\, \Big[ N(\tau ,{\vec \sigma}_1)\, \{
  \sigma^{\bar A}(\tau ,\vec \sigma ), {\cal H}(\tau ,{\vec
  \sigma}_1) \} + N_r(\tau ,{\vec \sigma}_1)\, \{  \sigma^{\bar A}(\tau ,\vec \sigma ),
  {\cal H}^r(\tau ,{\vec \sigma}_1) \}\Big] =\nonumber \\
  &=& {\bar \psi}^A(\tau ,\vec \sigma ) \approx 0.
  \label{5}
  \eea

\noindent Finally, ${\dot {\bar \psi}}^A(\tau ,\vec \sigma )
\approx 0$ determines the Dirac multipliers $\lambda^A(\tau ,\vec
\sigma )$.
\bigskip

In conclusion, the gauge fixings (\ref{4}) ({\it which break
general covariance}) constitute the crucial bridge that transforms
the {\it intrinsic pseudo-coordinates} into {\it true physical
individuating coordinates}. As a matter of fact, after going to
Dirac brackets, the point-events individuation is enforced in the
form of the {\it identity}

\beq
 \sigma^A \equiv {\bar \sigma}^{\bar A} = {\tilde F}^{\bar A}_{G}[
r^G_{\bar a}(\tau ,\vec \sigma ), \pi^G_{\bar a}(\tau , \vec
\sigma)] = F^{\bar A}[\Lambda^{(k)}_W(\tau ,\vec \sigma )]{|}_G.
 \label{6}
  \eeq

\medskip

In this {\it physical 4-coordinate grid}, the 4-metric, as well as
other  fundamental physical entities, like e.g. the space-time
interval $ds^{2}$ with its associated causal structure, and the
lapse and shift functions, depend entirely on the DO in that
gauge. Only on the solutions of Einstein's equations the
completely fixed gauge $G$ is equivalent to the fixation of a
definite 4-coordinate system $\sigma^A_G$. The gauge fixing
(\ref{4}) ensures that {\it on-shell} one gets $\sigma^A =
\sigma^A_G$. In this way we get a physical 4-coordinate grid on
the mathematical 4-manifold $M^4$ dynamically determined by
tensors over $M^4$ with a rule which is invariant under
${}_PDiff\, M^4$ but such that the functional form of the map {\it
$\sigma^A \mapsto \, physical\,\,\, 4-coordinates$} depends on the
complete chosen gauge $G$. This gauge-fixing makes the {\it
invariant pseudo-coordinates} into effective {\it individuating
fields} by forcing them to be {\it numerically} identical with
ordinary coordinates: in this way the individuating fields turn
the {\it mathematical} points of space-time into {\it physical
point-events}. What really individuates space-time points
physically are the very {\it degrees of freedom of the
gravitational field}. As a consequence,\- one can advance the {\it
ontological} claim that - physically - Einstein's vacuum
space-time is literally {\it identified} with the autonomous
physical degrees of freedom of the gravitational field, while the
specific functional form of the {\it invariant pseudo-coordinates}
matches these latter into the manifold's points. The introduction
of matter has the effect of modifying the Riemann and Weyl
tensors, namely the curvature of the 4-dimensional substratum, and
to allow {\it measuring} the gravitational field in a geometric
way for instance through effects like the geodesic deviation
equation. It is important to emphasize, however, that the addition
of {\it matter} does not modify the construction leading to the
individuation of point-events, rather it makes it {\it
conceptually more appealing}.
\bigskip

The gauge fixings (\ref{4}), (\ref{5}) induce a {\it
coordinate-dependent non-commutative Poisson bracket structure}
upon the {\it physical point-events} of space-time by means of the
associated Dirac brackets implying Eqs.(\ref{6}). More exactly,
on-shell, each coordinate system gets a well defined
non-commutative structure determined by the associated functions
${\tilde F}^{\bar A}_G(r^G_{\bar a}, \pi^G_{\bar a})$, for which
we have $\{ {\tilde F}^{\bar A}_G(r^G_{\bar a}(\tau , \vec \sigma
), \pi^G_{\bar a}(\tau ,\vec \sigma )), {\tilde F}^{\bar
B}_G(r^G_{\bar a}(\tau ,{\vec \sigma}_1), \pi^G_{\bar a}(\tau
,{\vec \sigma}_1)) \}^* \not= 0$. The physical implications of
this circumstance might deserve some attention in view of the
quantization of general relativity.

\bigskip

After this solution of the problem of the identification of the
point-events let us clarify the concept of {\it Bergmann's
observable} (BO) \cite{30a}. Bergmann's definition has various
facets, namely a {\it configurational} side having to do with
invariance under {\it passive} diffeomorphisms, an Hamiltonian
side having to do with Dirac's concept of observable, and the
property of {\it predictability} which is entangled with both
sides. According to Bergmann, (his) {\it observables} are passive
diffeomorphisms invariant quantities (PDIQ) "which can be
predicted uniquely from initial data", or "quantities that are
invariant under a coordinate transformation that leaves the
initial data unchanged". Bergmann says in addition that they are
further required to be gauge invariant, a statement that can only
be interpreted as implying that Bergmann's observables are
simultaneously DO. Yet, he offers no explicit demonstration of the
compatibility of this bundle of statements. The clarification of
this entanglement leads  to the proposal of a {\it main
conjecture} asserting: i) {\it the existence of special Dirac's
observables which are also Bergmann's observables}, as well as to
ii) {\it the existence of gauge variables that are coordinate
independent} (namely they behave like the tetradic scalar fields
of the Newman-Penrose formalism \cite{31a}).

\medskip

The Hamiltonian approach also allows to deduce something new
concerning the {\it overall role of gravitational and gauge
degrees of freedom}. Indeed, the distinction between gauge
variables and DO provided by the Shanmugadhasan  transformation
(\ref{3}), conjoined with the circumstance that the Hamiltonian
point of view brings naturally to a re-reading of geometrical
features in terms of the traditional concept of {\it force}, leads
to a by-product which should be added to the traditional wisdom of
the equivalence principle asserting the local impossibility of
distinguishing gravitational from inertial effects. Actually, the
isolation of the gauge arbitrariness from the true intrinsic
degrees of freedom of the gravitational field is instrumental to
understand and visualize which aspects of the local effects,
showing themselves on test matter, have a {\it genuine
gravitational origin} and which aspects depend solely upon the
choice of the (local) reference frame and could therefore even be
named {\it inertial} in analogy with their non-relativistic
Newtonian counterparts. Indeed, two main differences characterize
the issue of {\it inertial effects} in general relativity with
respect to the non-relativistic situation: the existence of {\it
autonomous degrees of freedom} of the gravitational field
independently of the presence of matter sources, on the one hand,
and the {\it local nature of the general-relativistic reference
systems}, on the other. Although the very definition of {\it
inertial forces} (and of {\it gravitational force} in general) is
rather arbitrary in general relativity, it appears natural to
characterize first of all as genuine gravitational effects those
which are directly correlated to the DO, while the gauge variables
appear to be correlated to the general relativistic counterparts
of Newtonian inertial effects. Another aspect of the Hamiltonian
connection "{\it gauge variables - inertial effects}" is related
to the 3+1 splitting of space-time required for the canonical
formalism. Each splitting is associated with a foliation of
space-time whose leaves are {\it Cauchy simultaneity} space-like
hyper-surfaces. While the field of unit normals to these surfaces
identifies a {\it surface-forming congruence of time-like
observers}, the field of the evolution vectors identifies a {\it
rotating congruence of time-like observers}. Since a variation of
the gauge variables modifies the foliation, the identification of
the two congruences of time-like observers is connected to the
fixation of the gauge, namely, on-shell, to the choice of
4-coordinates. Then a variation of gauge variables also modifies
the inertial effects.
\medskip

It is clear by now that a complete gauge fixing within canonical
gravity has the following implications: i) the choice of a unique
3+1 splitting with its associated foliation; ii) the choice of
well-defined congruences of time-like observers; iii) the {\it
on-shell} choice of a unique 4-coordinate system. In physical
terms this set of choices amount to choosing a {\it network of
intertwined and synchronized local laboratories made up with test
matter} (obviously up to a coherent choice of chrono-geometric
standards). This interpretation shows that, unlike in ordinary
gauge theories where the gauge variables are {\it inessential}
degrees of freedom, in general relativity they describe
generalized inertial effects.

\medskip

The only weakness of the previous distinction is that the
separation of the two autonomous degrees of freedom of the
gravitational field from the gauge variables is, as yet, a
coordinate (i.e. gauge) - dependent concept. The known examples of
pairs of conjugate DO are neither coordinate-independent (they are
not PDIQ) nor tensors. Bergmann asserts that the only known method
(at the time) to build BO is based on the existence of
Bergmann-Komar invariant pseudo-coordinates. A possible starting
point to attack the problem of the connection of DO with BO seems
to be a Hamiltonian re-formulation of the Newman-Penrose formalism
\cite{31a} (it contains only PDIQ) employing Hamiltonian
null-tetrads carried by the time-like observers of the congruence
orthogonal to the admissible space-like hyper-surfaces. This is
the source of the quoted {\it main conjecture} that special
Darboux bases for canonical gravity should exist in which the
inertial effects (gauge variables) are described by PDIQ while the
autonomous degrees of freedom (DO) are {\it also} BO.  Note that,
since Newman-Penrose PDIQ are tetradic quantities, the validity of
the conjecture would also eliminate the existing difference
between the observables for the gravitational field and the
observables for matter, built usually by means of the tetrads
associated to some time-like observer. Furthermore, this would
also provide a starting point for defining a metrology in general
relativity in a generally covariant way\footnote{Recall that this
is the main conceptual difference from the non-dynamical metrology
of special relativity}, replacing the empirical metrology
\cite{32a} used till now. It would also enable to replace by
dynamical matter the {\it test matter} of the axiomatic approach
\cite{33a} to measurement theory. This would constitute an
important advance, if we recall that all of the presentations of
gravitational waves and gravito-magnetism are till now
coordinate-dependent. Moreover, since no-one is able to solve the
super-hamiltonian constraint, it would be interesting to see how
it could be expressed in such a canonical basis. After all,
Ashtekar's approach started from a canonical transformation!

\bigskip

In Ref.\cite{8a} there is also a suggestion of how the physical
individuation of space-time points, introduced at the conceptual
level, could in principle be implemented with {\it a well-defined
empirical procedure, an experimental set-up and protocol for
positioning and orientation} based on the technology of the Global
Positioning System. This suggestion closes the {\it coordinative
circuit} of general relativity correlating the theoretical
construction with an empirical definition of space-time.

\bigskip

In conclusion the rest-frame instant form of metric and tetrad
gravity identifies a class of space-times where it is possible to
find an answer to all the interpretation problems of general
relativity. In them it is also possible to define a Hamiltonian
linearization in a completely fixed non-harmonic 3-orthogonal
gauge (the 3-metric is diagonal), which identifies a class of
linearized post-Minkowskian vacuum Einstein space-times
corresponding to {\it background-independent gravitational waves}
\cite{34a} (see the talk of De Pietri at this Conference). The
effect of the addition of matter (a relativistic perfect fluid) is
now under investigation. In presence of a perfect fluid this type
of background-independent linearization will make possible to
define a {\it weak field fast motion} approximation, to find the
form of the action-at-a-distance Newton and gravito-magnetic
potentials and of the Dirac observable (i.e. tidal) - fluid
interactions in this 4-coordinate system, without never making
post-Newtonian expansions, and finally to find the relativistic
quadrupole emission formula.
\bigskip

In generally covariant theories, the necessity of a physical
identification of point-events and the chrono-geometrical aspect
of the gravitational field (which teaches causality to all the
other fields) put the (graviton-like) physical degrees of freedom
of the gravitational field on a different level with respect to
photons, gluons... This seems to be in total contrast with all the
formulations on a background (like perturbative field theory and
string theory). As a consequence another important motivation for
looking for a canonical basis in which the gauge variables are
coordinate-independent and the DO are also BO, is to try to define
a new quantization scheme (respecting relativistic causality) for
canonical gravity, hopefully in a Fock space and not in
inequivalent Hilbert space like it happens in loop quantum
gravity. In a paper in preparation\cite{35a} this new quantization
scheme is defined and applied to get relativistic and
non-relativistic quantum mechanics in non-inertial frames in
absence of gravity (as an attempt to describe inertial effects in
a framework where they are no genuine tidal, i.e. DO, effects).

Till now there are two (nearly always inequivalent) families of
quantization schemes for systems with first class constraints:

i) first quantize all the canonical variables in a non physical
Hilbert space and then make the reduction with respect to the
gauge group arriving at the physical Hilbert space (usually a
quotient); in all the approaches (BRST, geometric, algebraic and
refined quantizations, deformations,...) the big problem is how to
determine the physical scalar product;

ii) first reduce and then quantize; here the problem is that
usually the classical reduced phase space is a highly
topologically non trivial manifold.

The idea behind the new quantization scheme is to arrive directly
to the physical Hilbert space by quantizing only the DO of the
system and treating the gauge variables as {\it c-numbers} (like
{\it time} in the time-dependent Schroedinger equation; the gauge
momenta become derivatives with respect to the gauge variables,
like the energy is replaced by the time derivative). In canonical
gravity this scheme would make sense only if the gauge variables
are coordinate-independent. There will be as many coupled
Schroedinger equations as gauge variables (plus eventually one
with the canonical Hamiltonian, when it is not vanishing) and the
wave function will depend on as many {\it times} (besides the
standard one) as gauge variables. Every line in this parameter
space will correspond to a gauge of the classical theory. If there
is an ordering such that the quantum constraints obtained with
this prescription (no ordering problem for the gauge variables!)
satisfy a commutator algebra with the constraints on the left of
the structure functions, then the coupled Schroedinger equations
will be formally integrable, the physical Schroedinger scalar
product (induced by the Schroedinger equations, i.e. by the
constraints) will not depend on the {\it times} (gauge
independence) and the propagation from an initial set of times to
a final one will not depend on the path in the parameter space
joining these two sets. Many topological properties of the
classical reduced phase space will be hidden in the properties in
large of the parameter space (the new quantization scheme is only
a local approximation). The first quantization of this type was
obtained many years ago, in the framework of relativistic particle
mechanics with first class constraints \cite{36a}, with the
quantization of the two-body DrozVincent-Todorov-Komar model with
an instantaneous action-at-a-distance potential: i) the two gauge
variables are the two times of the two particles; ii) the
quantization gives two coupled Klein-Gordon equations; iii) in
turn these equations led to the identification of four different
physical scalar products, one for each branch of the mass spectrum
(for non-equal masses).


\begin{references}

\bibitem{1a}L.Lusanna, {\it Towards a Unified Description of the Four
Interactions in Terms of Dirac-Bergmann Observables}, invited
contribution to the book {\it Quantum Field Theory: a 20th Century
Profile}, of the Indian National Science Academy, ed.A.N.Mitra,
forewards by F.J.Dyson (Hindustan Book Agency, New Delhi, 2000)
(hep-th/9907081).

\bibitem{2a}L.Lusanna, {\it The N- and 1-Time Classical
Descriptions of N-Body Relativistic Kinematics and the
Electromagnetic Interaction}, Int.J.Mod.Phys. {\bf A12}, 645
(1997).

\bibitem{3a}A.Einstein, {\it Die Grundlage der allgemeinin Relativit\"atstheorie},
Annalen der Physik {\bf 49}, 769 (1916); translated by W.Perrett
and G.B.Jeffrey, {\it The Foundations of of the General Theory of
Relativity}, in {\it The Principle of Relativity} (Dover, New
York, 1952), pp.117-118.

\bibitem{4a}J.Stachel, {\it Einstein's Search for General
Covariance, 1912-1915}, paper read at the Ninth International
Conference on General Relativity and Gravitation, Jena 1980;
published in {\it Einstein and the History of General Relativity},
Einstein Studies, Vol.1, eds. D.Howard and J.Stachel
(Birkh\"auser, Boston, 1985), pp.63-100.


\bibitem{5a}J.Norton, {\it How Einstein found his Field
Equations: 1912-1915},  {\it Historical Studies in the Physical
Sciences} {\bf 14}, 252 (1984); reprinted in {\it Einstein and the
History of General Relativity}, Einstein Studies, Vol. I, eds.
D.Howard and J.Stachel (Birkh\"auser, Boston, 1986)
p.101.\hfill\break
 {\it Coordinate and Covariance: Einstein's View of Space-Time and
 the Modern View}, Found.Phys. {\bf 19}, 1215 (1989).\hfill\break
 {\it The Physical Content of General Covariance}, in {\it Studies
 in the History of General Relativity}, Einstein Studies, Vol.3,
 eds. J.Eisenstaedt and A.Kox (Birkh\"auser, Boston,
 1992)p.281.\hfill\break
 {\it General Covariance and the Foundations of General
 Relativity: Eight Decades of Dispute}, Rep.Prog.Phys. {\it 56},
 791 (1993).\hfill\break
 {\it The Hole Argument}, Stanford Encyclopedia of Philosophy,
 1999: http://plato.stanford.edu/entries/spacetime-holearg/.

\bibitem{6a}J.Norton, {\it Einstein, the Hole Argument and the
Realiy of Space}, in {\it Measurement, Realism and Objectivity},
ed. J.Forge (Reidel, Dordrecht, 1987).

\bibitem{7a}J.Earman and J.Norton, {\it What Price Space-Time
Substantivalism? The Hole Story}, British Journal for the
Philosophy of Science, {\bf 38}, 515 (1987).

\bibitem{8a}L.Lusanna and M.Pauri, {\it General Covariance and the
Objectivity of Space-Time Point-Events: The Physical Role of
Gravitational and Gauge Degrees of Freedom in General Relativity}
(gr-qc/0301040).

\bibitem{9a} M.Pauri and M.Vallisneri, {\it Ephemeral Point-Events: is there a
Last Remnant of Physical Objectivity?}, essay for the 70th
birthday of R.Torretti, Dialogos {\bf 79}, 263 (2002)
(gr-qc/0203014).\hfill\break
 L.Lusanna, {\it Space-Time, General Covariance, Dirac-Bergmann
Observables and Non-Inertial Frames}, talk at the 25th Johns
Hopkins Workshop {\it 2001: A Relativistic Space-Time Odyssey},
(gr-qc/0205039), eds D.Dominici et al. (World Scientific,
Singapore, 2003).

\bibitem{10a}P.G.Bergmann and A.Komar, {\it The Coordinate Group
Symmetries of General Relativity}, Int.J.Theor.Phys. {\bf 5}, 15
(1972).

\bibitem{11a}H.Weyl, {\it Groups, Klein's Erlangen Program.
Quantities}, ch.I, sec.4 of  {\it The Classical Groups, their
Invariants and Representations}, 2nd ed., (Princeton University,
Princeton, 1946), pp.13-23.

\bibitem{12a}J.Stachel, {\it How Einstein Discovered General
Relativity: A Historical Tale with Some Contemporary Morals}, in
{\it General Relativity and Gravitation}, ed. M.A.H. MacCallum
(Cambridge University Press, Cambridge, 1986) p.200.

\bibitem{13a}J.Stachel, {\it The Meaning of General Covariance}, in
{\it Philosophical Problems of the Internal and External Worlds},
Essays in the Philosophy of A.Gr\"unbaum, eds. J.Earman,
A.I.Janis, G.J.Massey and N.Rescher (Pittsburgh Univ. Press,
Pittsburgh, 1993).\hfill\break
 {\it How Einstein Discovered
General Relativity: a Historical Tale with Some Contemporary
Morals}, in Proc. GR11, ed. M.A.H.MacCallum (Cambridge Univ.Press,
Cambridge, 1987).


\bibitem{14a}P.G.Bergmann and A.Komar, {\it Poisson Brackets between
Locally Defined Observables in General Relativity},
Phys.Rev.Letters {\bf 4}, 432 (1960).

\bibitem{15a}H.Friedrich and A.Rendall, {\it The Cauchy Problem for
Einstein Equations}, in {\it Einstein's Field Equations and their
Physical Interpretation}, ed. B.G.Schmidt (Springer, Berlin, 2000)
(gr-qc/0002074).\hfill\break
 A.Rendall, {\it Local and Global Existence Theorems for the
 Einstein Equations}, Online journal Living Reviews in Relativity
 {\bf 1}, n. 4 (1998) and {\bf 3}, n. 1 (2000) (gr-qc/0001008).


\bibitem{16a}S.Shanmugadhasan, {\it Canonical Formalism for Degenerate
Lagrangians}, J.Math.Phys. {\bf 14}, 677 (1973).\hfill\break
 L.Lusanna, {\it The Shanmugadhasan Canonical Transformation, Function Groups
 and the Second Noether Theorem}, Int.J.Mod.Phys. {\bf A8}, 4193 (1993).

\bibitem{17a}P.A.M.Dirac, {\it Lectures on
Quantum Mechanics}, Belfer Graduate School of Science, Monographs
Series (Yeshiva University, New York, N.Y., 1964).

\bibitem{18a}R.Arnowitt, S.Deser and C.W.Misner, {\it Canonical Variables
for General Relativity}, Phys.Rev. {\bf 117}, 1595 (1960).
 \hfill\break
 {\it The Dynamics of General Relativity}, in {\it Gravitation: an
Introduction to Current Research}, ch. 7, ed.L.Witten (Wiley, New
York, 1962).



\bibitem{19a}L.Lusanna, {\it The Rest-Frame Instant Form of Metric
Gravity}, Gen.Rel.Grav. {\bf 33}, 1579 (2001)
(gr-qc/0101048).\hfill\break
 L.Lusanna and S.Russo, {\it A New Parametrization for Tetrad
Gravity}, Gen.Rel.Grav. {\bf 34}, 189
(2002)(gr-qc/0102074).\hfill\break
 R.De Pietri, L.Lusanna, L.Martucci and S.Russo, {\it Dirac's
Observables for the Rest-Frame Instant Form of Tetrad Gravity in a
Completely Fixed 3-Orthogonal Gauge}, Gen.Rel.Grav. {\bf 34}, 877
(2002) (gr-qc/0105084).

\bibitem{20a}D.Christodoulou and S.Klainerman, {\it The Global Nonlinear
Stability of the Minkowski Space} (Princeton, Princeton, 1993).

\bibitem{21a}J.Isenberg and J.E.Marsden, {\it The York Map is a Canonical
Transformation}, J.Geom.Phys. {\bf 1}, 85 (1984).

\bibitem{22a}L.Lusanna, {\it Classical Yang-Mills Theory with
Fermions}, {\it I) General Properties of a System with
Constrants}, Int.J.Mod.Phys. {\bf A10}, 3531 (1995); {\it II)
Dirac's Observables}, Int.J.Mod.Phys. {\bf A10}, 3675 (1995).


\bibitem{23a}A.Lichnerowicz, {\it Propagateurs, Commutateurs et
Anticommutateurs en Relativite Generale}, in Les Houches 1963,
{\it Relativity, Groups and Topology}, eds. C.DeWitt and B.DeWitt
(Gordon and Breach, New York, 1964).\hfill\break
 C.Moreno, {\it On the Spaces of Positive and Negative Frequency
 Solutions of the Klein-Gordon Equation in Curved Space-Times},
 Rep.Math.Phys. {\bf 17}, 333 (1980).


\bibitem{24a}N.D.Birrell and P.C.W.Davies, {\it Quantum Fields
in Curved Space} (Cambridge University Press, Cambridge, 1982).
\hfill\break
 P.C.W.Davis, {\it Particles do not Exist}, in {\it Quantum Theory
 of Gravity. Essays in Honor of the 60th Birthday of Bryce DeWitt.}, ed.
 S.Christensen (Hilger, Bristol, 1984).




\bibitem{25a}B.DeWitt, {\it Quantum Theory of Gravity}, {\it I) The Canonical
Theory}, Phys.Rev. {\bf 160}, 1113 (1967), {\it II) The Manifestly
Covariant Theory}, {\bf 162}, 1195 (1967).



\bibitem{26a}Y.Choquet-Bruhat, A.Fischer and J.E.Marsden, {\it Maximal
Hyper-Surfaces and Positivity of Mass}, LXVII E.Fermi Summer
School of Physics {\it Isolated Gravitating Systems in General
Relativity}, ed. J.Ehlers (North-Holland, Amsterdam, 1979).




\bibitem{27a}G.Belot and J.Earman, {\it From Metaphysics to
Physics}, in {\it From Physics to Philosophy}, eds J.Butterfield
and C.Pagonis (Cambridge University Press, Cambridge, 1999) p.166.
 \hfill\break
{\it Pre-Socratic Quantum Gravity}, in {\it Physics Meets
Philosophy at the Planck Scale. Contemporary Theories in Quantum
Gravity}, ed. C.Callender (Cambridge University Press, Cambridge,
2001).\hfill\break
 J.Earman, {\it Thoroughly Modern McTaggart or what McTaggart would
have said if He had read the General Theory of Relativity},
Philosophers' Imprint {\bf 2}, No.3 August 2002
(http://www.philosophersimprint.org/002003/).



\bibitem{28a}P.G.Bergmann, {\it The General Theory of Relativity}, in {\it Handbuch
der Physik}, Vol. IV, {\it Principles of Electrodynamics and
Relativity}, ed. S.Flugge (Springer, Berlin, 1962) pp.247-272.


\bibitem{29a}A.Komar, {\it Construction of a Complete Set of Independent
Observables in the General Theory of Relativity}, Phys.Rev. {\bf
111}, 1182 (1958).


\bibitem{30a}P.G.Bergmann, {\it Observables in General Relativity},
Rev.Mod.Phys. {\bf 33}, 510 (1961).

\bibitem{31a}J.Stewart, {\it Advanced General Relativity} (Cambridge Univ.
Press, Cambridge, 1993).

\bibitem{32a}M.H.Soffel, {\it Relativity in Astrometry, Celestial Mechanics
and Geodesy} (Springer, Berlin, 1989).

\bibitem{33a}J.Ehlers, F.A.E.Pirani and A.Schild, {\it The Geometry of
Free-Fall and Light Propagation} in {\it General Relativity,
Papers in Honor of J.L.Synge}, ed. L.O'Raifeartaigh (Oxford
Univ.Press, London, 1972).

\bibitem{34a}J.Agresti, R.DePietri, L.Lusanna and L.Martucci, {\it
Hamiltonian Linearization of the Rest-Frame Instant Form of Tetrad
Gravity in a Completely Fixed 3-Orthogonal Gauge: a Radiation
Gauge for Background-Independent Gravitational Waves in a
Post-Minkowskian Einstein Space-Time} (gr-qc/0302084) and {\it
Background Independent Gravitational Waves} (gr-qc/0302085).

\bibitem{35a}D.Alba and L.Lusanna, {\it From Parametrized Minkowski
Theories to Quantum Mechanics in Non-Inertial Frames in Absence of
Gravity: I) Relativistic and Non-Relativistic Particles}, in
preparation.

\bibitem{36a}G.Longhi and L.Lusanna, {\it Bound-State Solutions,
Invariant Scalar Products and Conserved Currents for a class of
Two-Body Relativistic Systems}, Phys.Rev. D34, 3707 (1986).




\end{references}
\end{document}